\documentclass[aps,pra,showpacs,twocolumn,floatfix,10pt]{revtex4-1}
\usepackage{graphicx}
\usepackage{amssymb}
\usepackage{subfigure}
\usepackage{amsmath}
\usepackage{color}

\begin{document}
\title{Radio-frequency-modulated Rydberg states in a vapor cell}
\author{S.~A.~Miller$^1$}
\author{D.~A.~Anderson$^2$}
\author{G.~Raithel$^1$}
\affiliation{1. Department of Physics, University of Michigan, Ann Arbor, MI 48109}
\affiliation{2. Rydberg Technologies, LLC, Ann Arbor, MI 48104}

\date{January 12, 2016}

\begin{abstract}
  We measure strong radio-frequency (RF) electric fields using rubidium Rydberg atoms prepared in a room-temperature vapor cell as field sensors.  Electromagnetically induced transparency is employed as an optical readout.  We RF-modulate the 60$S_{1/2}$ and 58$D_{5/2}$ Rydberg states with 50~MHz and 100~MHz fields, respectively.  For weak to moderate RF fields, the Rydberg levels become Stark-shifted, and sidebands appear at even multiples of the driving frequency. In high fields, the adjacent hydrogenic manifold begins to intersect the shifted levels, providing rich spectroscopic structure suitable for precision field measurements. A quantitative description of strong-field level modulation and mixing of $S$ and $D$ states with hydrogenic states is provided by Floquet theory. Additionally, we estimate the shielding of DC electric fields in the interior of the glass vapor cell.
\end{abstract}


\maketitle

\section{Introduction}
Recently there has been a growing interest in using electromagnetically induced transparency (EIT) with Rydberg atoms to study atomic spectra in microwave fields and, conversely, to employ the atomic response to measure properties of the field. Developments include measurements of weak and strong microwave electric fields (ranging from $\sim$1~mV/m~\cite{ShafferSensitivity} up to 230~V/m~\cite{TwoPhoton,HighPower}), microwave polarization measurements~\cite{Sedlacek.2013}, and subwavelength imaging~\cite{SubwavelengthImaging,Subwavelength_Shaffer}. These efforts are further motivated by a broader initiative towards establishing atomic measurement standards for field quantities~\cite{ShafferSensitivity, Budker.2007, Sheng.2013}.  A significant appeal of the demonstrated technique for measurements of microwave fields is that it offers a wide dynamic range in both field strength and frequency, from tens of GHz up to the sub-THz regime~\cite{Broadband,MMwaves}. Another noteworthy benefit is that Rydberg EIT experiments can be performed in a room-temperature vapor cell and without laser cooling or particle detection methods, resulting in a significantly simplified system.

When atoms are exposed to electric fields, their energy levels are Stark shifted. For hydrogenic Rydberg states (states with high angular momentum $\ell$ and very small quantum defects) in DC electric fields, the shift lifts their degeneracy, resulting in a spectral map that resembles a fan~\cite{Kleppner1979, Gallagher}, which has also recently been measured in a room-temperature cell using EIT~\cite{HighDC}. As the electric field increases, the red- or blue-shifted states of the fan eventually intersect with the Rydberg states that are low in angular momentum (in Rb $\ell\leq2$) and are initially outside the fan of hydrogenic states. The intersections are characterized by sequences of level crossings and state mixing of high- and low-$\ell$ Rydberg states. Rydberg atoms in radio-frequency (RF) fields exhibit a similar behavior, with added complexity due to the presence of dense sets of RF-dressed states.

When weak RF fields are applied to the system, the Rydberg levels experience AC Stark shifts proportional to the cycle average of the RF intensity. As the RF field increases, sidebands begin to emerge. In the presence of a harmonic AC field, only even sidebands appear (i.e. sidebands separated by an even multiple of the RF driving frequency). A DC field offset produces odd sidebands (i.e. sidebands separated by an odd multiple of the RF driving frequency). In previous work on the effects of AC and DC fields in Rydberg atoms, low-order perturbation theory was employed to describe the behavior of the atoms~\cite{NJP_Adams, SidebandShifts}. There, the response of the Rydberg level to the AC field is accounted for using the quadratic Stark effect and a first-order correction to the wave function.  A prominent feature of these low-field spectral maps are dropouts of the RF bands, which follow from Bessel-function zeros in the line strength calculated for the bands (see Eqn.~6 in~\cite{NJP_Adams}, for example). A non-perturbative Floquet treatment has also been employed to describe the effects of AC and DC fields in hydrogenic Rydberg states of sodium in the limit of weak field amplitudes~\cite{Floquet_Hydrogen}.

In this paper, we use room-temperature EIT to spectroscopically investigate 60$S_{1/2}$ and 58$D_{5/2}$ Rydberg states modulated with 50 and 100~MHz RF radiation, respectively, over a wide range of RF power. We investigate the $S$ and $D$ states in strong fields where the level shifts are no longer quadratic and where they intersect with the fan of hydrogenic states. The level shifts and excitation rates observed in our strong-field experimental spectra are modeled using a non-perturbative Floquet treatment, which accounts for high-order state mixing, as well as multi-RF-photon absorption and emission. We obtain excellent agreement between our experimental and calculated spectral maps, validating the use of the Floquet model to obtain a quantitative understanding of the atomic response to strong RF fields and allowing us to extract quantitative information on the applied RF field.  By this approach, we measure an RF electric field amplitude to within an uncertainty of $\pm$0.35$\%$, almost an order of magnitude better than the uncertainty of standard RF electric field calibration probes~\cite{Hill.1990,Matloubi.1993}. In addition, we briefly discuss the DC shielding effect of the glass vapor cell, which is of direct relevance to low-frequency field measurements by this experimental approach.

\section{Experimental Setup}
\begin{figure}
\includegraphics[width=8.5cm]{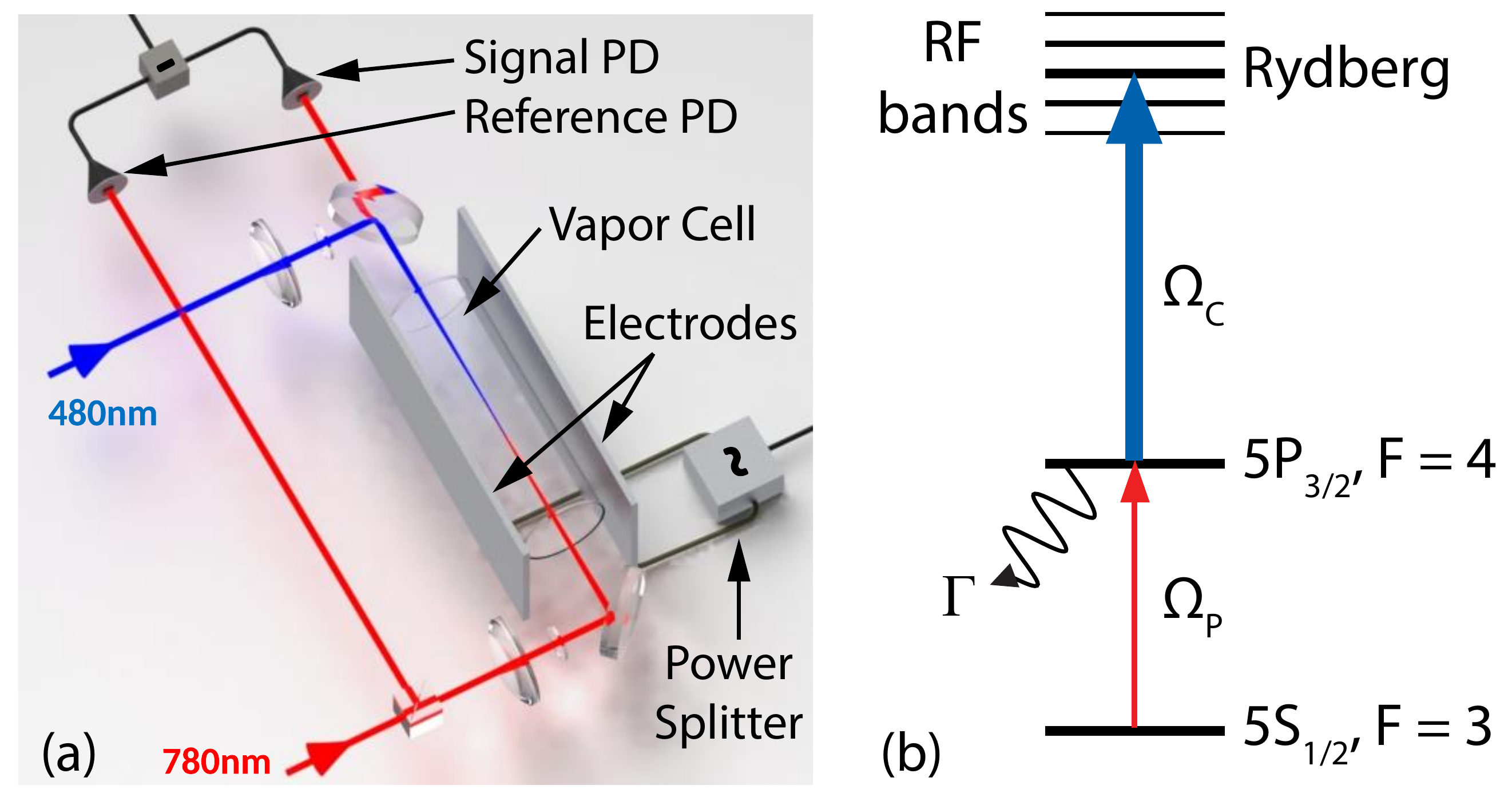}
\caption[caption]{(Color online) (a) Experimental setup including a vapor cell, 780~nm probe and reference beams (through and outside the vapor cell, respectively), 480~nm coupling beam, photodiodes for detection, RF power splitter (with a 180$^\circ$ phase difference between the outputs), and electrode plates. (b) Energy level diagram of the $^{85}$Rb Rydberg-EIT transitions, where the $\Omega_P$ and $\Omega_C$ are the probe and coupling Rabi frequencies, respectively, and $\Gamma$ is the intermediate-state decay rate.}
\label{experimental_setup}
\end{figure}

The experimental setup and energy level diagram are shown in Fig.~\ref{experimental_setup}(a) and (b), respectively. The setup consists of a cylindrical room-temperature rubidium vapor cell that is 75~mm long and 25~mm in diameter. The two lasers used have wavelengths 780~nm and 480~nm. The 780~nm probe beam is split into two paths: one through the vapor cell for EIT and one outside the cell. The two beams are incident on signal and reference photodiodes, respectively, and the difference signal is sent to an oscilloscope where it is recorded. The differential method reduces noise due to laser power fluctuations. The beam passing through the vapor cell is focused at the center of the cell to a full-width-at-half-maximum (FWHM) of 70~$\mu$m with a power of $\approx$~5~$\mu$W. The probe laser is locked to the $|5S_{1/2}, F=3>$ $\rightarrow$ $|5P_{3/2}, F=4>$ transition of $^{85}$Rb. The 480~nm coupling beam is overlapped with the 780~nm probe beam inside the cell, and focused to a FWHM of 84~$\mu$m with a power of $\approx$~45~mW. The coupling laser frequency is scanned over a range of about 1~GHz over the $|5P_{3/2}, F=4>$ $\rightarrow$ $|Rydberg>$ transition. A weak reference beam derived from the coupling laser is sent through a Fabry-Perot cavity [not shown in Fig.~\ref{experimental_setup}(a)], whose transmission peaks are used as frequency reference points for the EIT spectra to correct for small frequency drifts of the coupling laser.  As indicated in Fig.~\ref{experimental_setup}(b), the relevant frequencies characterizing the EIT system~\cite{Berman} are: the coupler Rabi frequency ($\Omega_{C, 60S} \approx 2\pi \times 4$~MHz, $\Omega_{C, 58D} \approx 2\pi \times 8$~MHz), the probe Rabi frequency ($\Omega_P \approx 2\pi \times 26$~MHz), the 5$P_{3/2}$ decay rate ($\Gamma = 2\pi \times 6$~MHz), and the laser linewidths ($\lesssim$~1~MHz). The reduction in absorption (i.e. the EIT transparency), measured as a function of the scanned coupler-laser frequency, reveals the $|5P_{3/2}, F=4>$ $\rightarrow$ $|Rydberg>$ excitation spectrum of the atoms under presence of the applied RF field. We note that a high probe Rabi frequency was chosen in order to obtain high signal strength, achieved at the expense of an increased EIT linewidth. In the present work, signal strength was more important than reaching optimal spectral resolution.

The rubidium vapor cell is suspended between two aluminum plate electrodes that are separated by 2.8~cm, slightly more than the cell diameter. The RF signal is applied to the plates, generating a RF field at the location of the atoms in the cell. The RF is supplied by a signal generator, the output of which is passed through a 40~dB amplifier and split by a 180$^\circ$ power splitter. For all experiments, the optical beams are linearly polarized, with the polarizations parallel to the RF electric-field direction.

\section{Experimental Results}

\begin{figure}
\includegraphics[width=7cm]{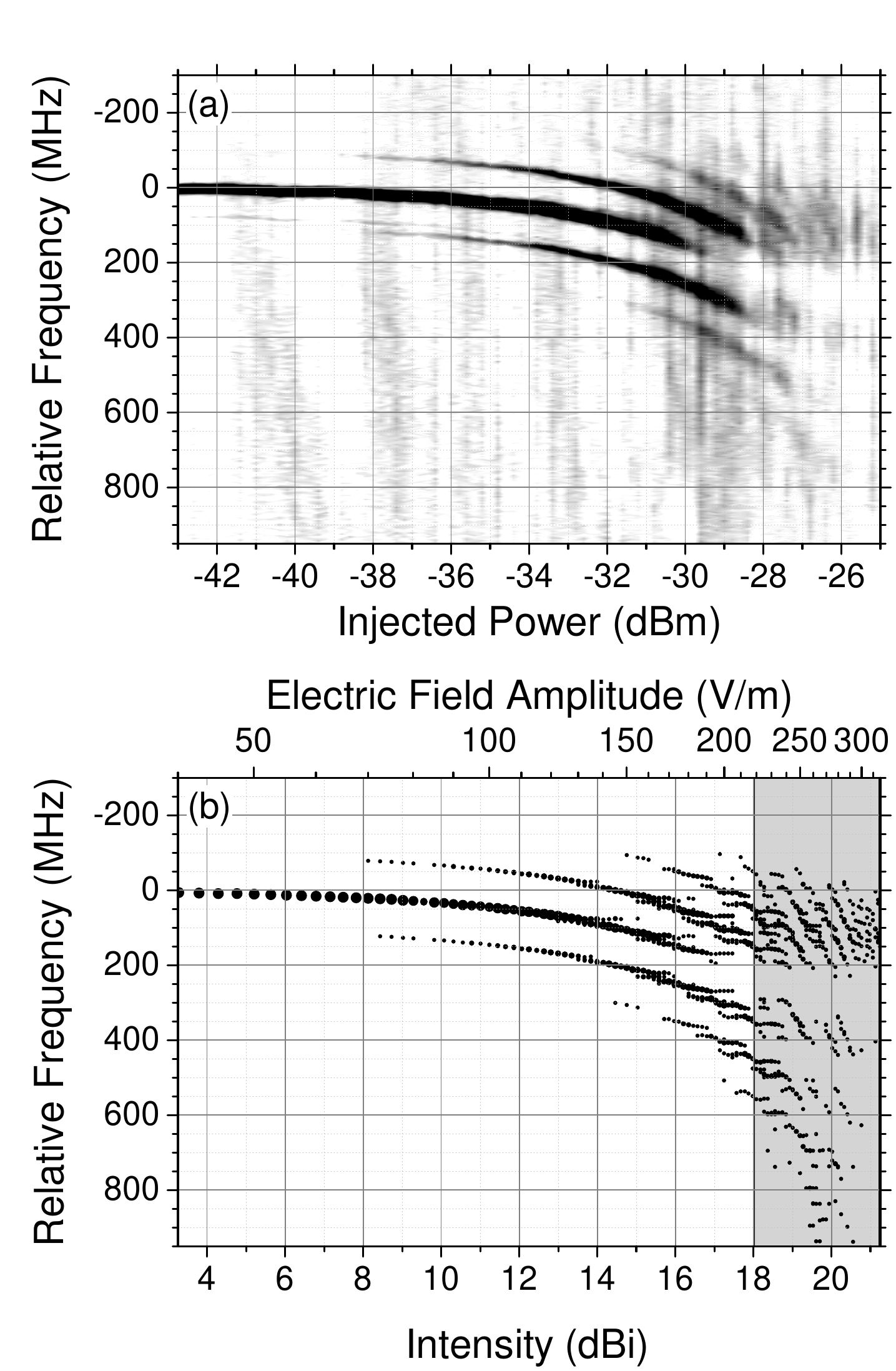}
\caption[caption]{Experimental (a) and calculated (b) spectra of the 60$S_{1/2}$ state modulated by 50~MHz RF fields. (a) Scale ranges from 0 (white) to $\geq$ 0.01 (black) in arbitrary units. (b) The area of each dot is proportional to the calculated excitation rate at that point. The dBi-value for intensity $I$ is~$10\log_{10}[I/(1~$W/m$^2)]$.  The gray shaded region indicates the domain in which the 60$S_{1/2}$ level mixes with the fan of hydrogenic states.}
\label{figure60S_50MHz}
\end{figure}

\begin{figure}
\includegraphics[width=8.5cm]{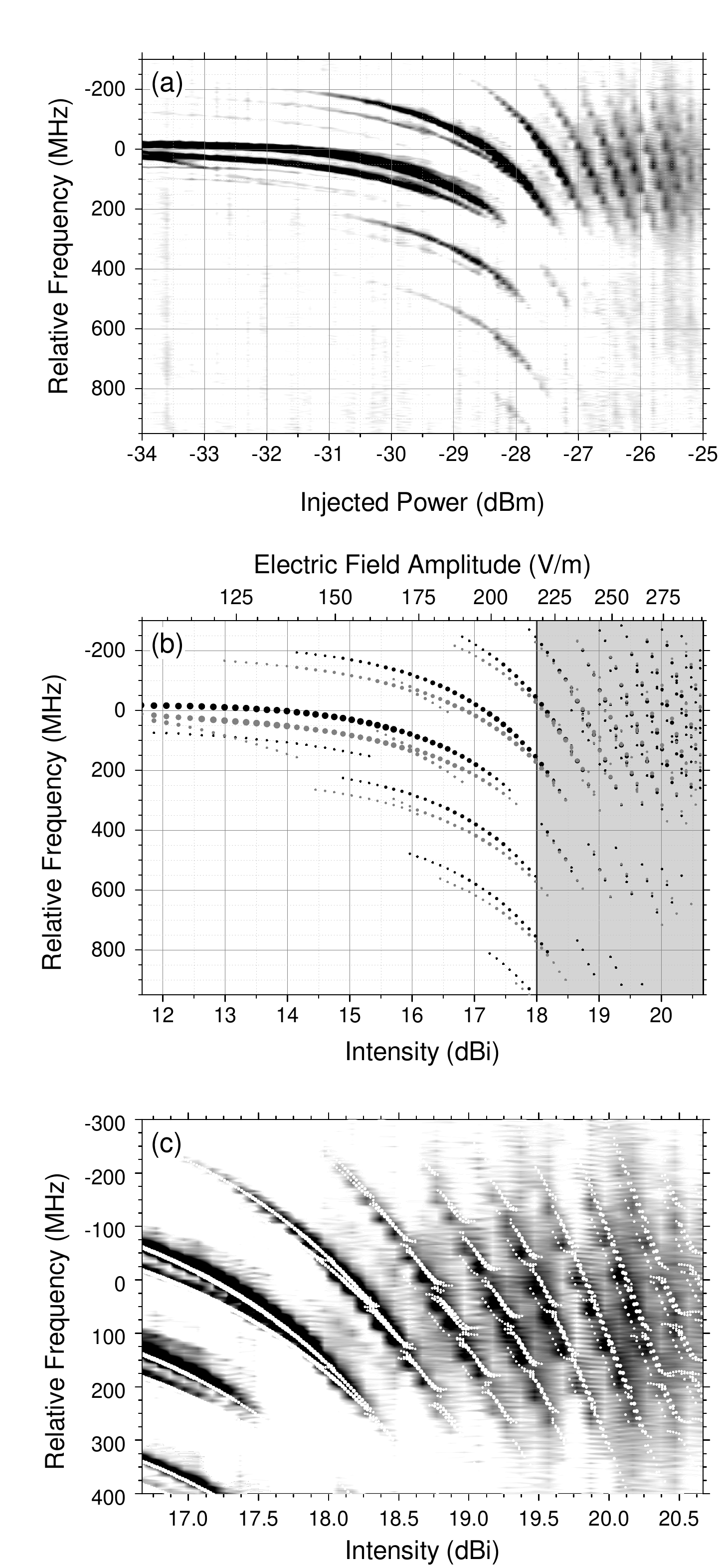}
\caption[caption]{Experimental (a) and calculated (b) spectra of the 58$D$ state driven by 100~MHz RF fields. (a) Scale ranges from 0 (white) to $\geq$0.01 (black) in arbitrary units. (b) Black dots are $m_j$~=~1/2 and gray dots are $m_j$~=~3/2. The gray shaded region indicates the domain in which the 58$D$ levels mix with the fan of hydrogenic states. (c) Enlarged section of top figure overlaid with a higher-resolution calculated spectrum. Scale ranges from 0 (white) to $\geq$0.007 (black) in arbitrary units. For the calculated spectra, the area of each dot is proportional to the calculated excitation rate at that point. The dBi-value for intensity $I$ is~$10\log_{10}[I/(1~$W/m$^2)]$.}
\label{figure_58D_triple}
\end{figure}

First, we investigate the 60$S_{1/2}$ Rydberg state modulated by 50~MHz RF fields. The power injected into the amplifier is varied from -43 to -25~dBm in steps of 0.2~dBm. The resulting experimental spectrum is shown in Fig.~\ref{figure60S_50MHz}(a). As the RF power is increased, we see that the first set of RF sidebands emerges at about $-38$~dBm. These sidebands are of second order, with a separation from the main AC-Stark-shifted Rydberg line of 100~MHz. The fourth order harmonics appear at about -31~dBm. As expected from the harmonic character of the AC drive signal, no odd harmonics are observed~\cite{NJP_Adams}. When an injection power of -28~dBm is reached, the modulated S line begins to mix with the hydrogenic states [see corresponding gray shaded box of Fig.~\ref{figure60S_50MHz}(b)]. In the mixing regime the signal strength is significantly reduced, because the 5$P_{3/2}$-60$S_{1/2}$ coupler-transition oscillator strength spreads over an increasing number of RF-dressed Rydberg levels. Since the 5$P_{3/2}$-60$S_{1/2}$ oscillator strength is relatively small, the spectroscopic signal drops below the experimental sensitivity limit at about -26~dBm.

To study a regime of stronger mixing, we turn to the nearby 58$D_{5/2}$ level. Since the $D$ ($\ell =2$) states have higher angular momentum than $S$ ($\ell =0$) states, they couple with the hydrogenic lines in first order, leading to stronger mixing in the 58$D_{5/2}$ case than in the 60$S_{1/2}$ case. Furthermore, the 58$D_{5/2}$ level is more strongly coupled to the 5$P_{3/2}$ intermediate level than the 60$S_{1/2}$ level is, which is desirable for an improved signal-to-noise ratio. To get a larger separation between the RF-dressed Rydberg levels, we drive the 58$D_{5/2}$ state with 100~MHz. Figure~\ref{figure_58D_triple}(a) shows the experimental spectrum of the modulated 58$D_{5/2}$ state for RF powers ranging from -34 to -25~dBm in steps of 0.1~dBm. The two dominant lines at low field are the 58$D_{5/2}$, $m_j$=1/2 and 3/2 components. The weaker 58$D_{3/2}$ fine structure component also appears in the spectrum (for 58$D$ the fine structure splitting is 58.7~MHz).  In the RF-dressed sidebands, both $m_j$ components of the 58$D_{5/2}$ state are present as well. At high powers, there are multiple distinguishable lines originating from the fan of hydrogenic states [see corresponding gray shaded box of Fig.~\ref{figure_58D_triple}(b)]. The slopes of these lines are very steep, providing excellent features for calibrating the field produced by the RF source.

In Fig.~\ref{figure_58D_triple}(a) and (c) at higher RF powers, the experimental signal appears to be spotted. This is an artifact resulting from the utilized 0.1~dBm step size of the signal generator. The spots are not clearly visible in Fig.~\ref{figure60S_50MHz}(a) due to the lower 5$P_{3/2}$-60$S_{1/2}$ oscillator strength and the resultant lower signal strength of the 60$S_{1/2}$ spectra.

\section{Floquet Analysis}
We classify the spectral maps in Figs.~\ref{figure60S_50MHz}(a) and \ref{figure_58D_triple}(a) into two regimes: low field ($<$~-28~dBm), where sidebands are present at equal intervals, and high field ($>$~-28~dBm), where the spectral lines are denser due to mixing with hydrogenic states. Moreover, both maps exhibit only even harmonics, indicating the absence of a DC field. The line strength for the 0$^{th}$ order harmonic drops to zero at about -28~dBm and reappears at higher RF power. In the high field regime, where the $S$ or $D$ Rydberg states RF-couple with the fan of hydrogenic states, the perturbative model developed in~\cite{NJP_Adams, SidebandShifts} is inadequate. To accurately describe energy levels and signal strengths at high RF powers, where higher-order RF couplings are prevalent, we use a Floquet treatment to provide an exact, non-perturbative description of the atomic response to harmonic AC fields.  Here, we include couplings between many Rydberg states, while allowing for the absorption or emission of an arbitrary number of RF photons during the optical excitation. A similar treatment is used in Ref.~\cite{Floquet}. The implemented Floquet method has previously been established as a suitable means to model Rydberg-atom microwave spectra in vapor cell experiments~\cite{TwoPhoton, HighPower}.

Relevant results of the theory include the coupling-laser frequencies, $\omega_{\nu, N}$, at which resonances are observed, as well as their relative excitation rates, $S_{\nu, N}$,
\begin{eqnarray}
\hbar \omega_{\nu, N} & = & W_\nu + N \hbar \omega_{RF} \nonumber  \\
S_{\nu, N} & = & (e F_L/ \hbar)^2 \left| \sum_k \tilde{C}_{\nu,k,N} \, {\hat{\bf{\epsilon}}} \cdot \langle k \vert \hat{\bf{r}} \vert 5P_{3/2}, m_j \rangle \right|^2, \quad
\label{eq:rates}
\end{eqnarray}
where $F_L$ is the amplitude of the laser electric field, ${\hat{\bf{\epsilon}}}$ is the laser polarization vector, and $\langle k \vert \hat{\bf{r}} \vert 5P_{3/2}, m_j \rangle$ are the electric-dipole matrix elements of the Rydberg basis states $\vert k \rangle = \vert n, \ell, j, m_j \rangle$ with $\vert 5P_{3/2}, m_j \rangle$.  Each Floquet level (energy $W_\nu$) carries RF-dressed sidebands in the spectrum, which are associated with the exchange of a number of $N$ RF photons during the excitation. The required Fourier coefficients $\tilde{C}_{\nu,k,N}$ follow from the time-dependent Floquet wave-packets, $\vert \Psi_{\nu}(t) \rangle$,  and exact time-periodic functions, $C_{\nu,k}(t)$, according to
\begin{eqnarray}
\vert \Psi_{\nu}(t) \rangle & = & {\rm{e}}^{-iW_{\nu}t/\hbar} \sum_k C_{\nu,k}(t)\vert k\rangle \nonumber \\
~                           & = & {\rm{e}}^{-iW_{\nu}t/\hbar} \sum_{k}\sum^{\infty}_{N=-\infty} \tilde{C}_{\nu,k,N}{\rm{e}}^{-i N \omega_{RF} t}\vert k\rangle, \nonumber \\
\tilde{C}_{\nu,k,N}&=&\frac{1}{T}\int_{0}^{T} C_{\nu,k}(t){\rm{e}}^{iN\omega_{RF}t}dt,  \quad
\label{eq:rates2}
\end{eqnarray}
with RF-field period $T$. For more details, see~\cite{HighPower}.

In the implementation of the theory, the basis $\{ \vert k \rangle \}$ must be chosen large enough to cover important couplings and small enough to allow for completion of the calculation within a reasonable time. Since the theory has to accurately describe mixing with hydrogenic states, the basis needs to include all possible $\ell$ and $j$-values, which are $\ell=0, ..., n-1$ and $j=\ell \pm 1/2$ (for $\ell=0$ only $j=1/2$). Also, since the fields are $\pi$-polarized, calculations are done separately for the allowed (fixed) $m_j$-values. Since the coupler-laser excitation originates from the $5P_{3/2}$ level, the accessible $m_j$-values are $\vert m_j \vert = 1/2, 3/2$ (the sign is irrelevant). For the present case, we have found that the basis can be limited to states with energies that correspond to effective principal quantum numbers $55.05 \le n_{\rm{eff}} \le 57.95$.

The implementation of the theory requires the integration of the time evolution operator through one RF cycle. The time step size $\Delta t$ must be chosen small enough that the relation $\vert W_k - W_{k'} \vert \Delta t / \hbar \le 2 \pi$ holds for all energy differences $W_k - W_{k'}$ for basis states $\vert k \rangle $ and $ \vert k' \rangle$. In the present case, 2048 to 4192 time steps are appropriate. When violating this condition, basis states with energies far away from 60$S_{1/2}$ and 58$D_{5/2}$ (our states of interest) generate aliases in the investigated region, resulting in wrong Floquet spectra.

Spectra calculated using Floquet theory are shown in Fig.~\ref{figure60S_50MHz}(b) and Fig~\ref{figure_58D_triple}(b) for the 60$S_{1/2}$ and 58$D_{5/2}$ states, respectively. Each laser frequency $\omega_{\nu, N}$ where a Floquet level is excited by an optical and $N$ RF photons is represented as a spot with an area proportional to the excitation rate of that level. The $x$-axis is the RF intensity expressed in units of dBi~=~10log$_{10}$(I/I$_0$), where I$_0$~=~1~W/m$^2$.

\section{Analysis}

Experimental and calculated spectra in Figs.~\ref{figure60S_50MHz} and \ref{figure_58D_triple} are in excellent quantitative agreement. Comparing experiment and theory, we find that we reach maximum field strengths of 317~V/m for 60$S_{1/2}$ and 296~V/m for 58$D_{5/2}$. Although both experimental spectra are obtained up to the same injected RF power level of -25~dBm, applied at the input of the 40~dB power amplifier, the maximum fields reached differ by 20~V/m.  This is attributed to a slight frequency dependence in the RF transmission line and generator performance (measured separately and not presented here).

We are able to measure the field (power) to within $\pm$0.88$\%$ ($\pm$1.75$\%$) for the 60$S_{1/2}$ state and to within $\pm$0.35$\%$ ($\pm$0.70$\%$) for the 58$D_{5/2}$ state. This level of precision is about a factor of ten better than that achieved by traditional dipole probes used as standard references for RF field calibration~\cite{Hill.1990,Matloubi.1993}. The level of precision achievable by our approach is related to the number of distinct features in the spectra and their associated dynamic dipole moments (states with larger dynamic dipole moments enable higher precision because they provide a tighter bound on the uncertainty in the alignment of experimental and calculated spectra). In spectral maps for strong RF fields, such as those presented here, high measurement precision can be readily obtained as a result of the high density of spectral features in the RF-modulated hydrogenic fan and the large dipole moments of the hydrogenic states.

In the case of the 60$S_{1/2}$ spectrum, the field measurement is obtained by matching the observed and calculated dropouts of the $N=0$ band of the 60$S_{1/2}$-line as well as the locations at which the $\vert N\vert=2$ and $\vert N\vert=4$ bands develop large negative slopes (equivalent to large AC dipole moments) due to mixing with hydrogenic levels. The observed dropout of the $N=0$ band is centered at 18.0~dBi, corresponding to an RF-field amplitude of $E_{RF}=218$~V/m. In comparison, the low-field perturbative model~\cite{NJP_Adams}, which applies to isolated levels with fixed DC-polarizability $\alpha$, predicts the first dropout of the $N=0$ band when $\alpha E_{RF}^2 /(8 \hbar \omega_{RF})=2.40$ (where the numerical value 2.40 is the first root of the first-kind Bessel function $J_0$). The Rb 60$S_{1/2}$ polarizability $\alpha=18.2$~kHz/(V/m)$^2$, which means that for a 50~MHz modulation field the first dropout of the $N=0$ band according to the low-field model would be at $E_{RF}=231$~V/m. The fact that the dropout occurs at a slightly smaller field can be traced back to the increase of the polarizability of 60$S_{1/2}$ at higher fields due to level repulsion from the hydrogenic states.

The 58$D_{5/2}$ spectrum has more features and higher overall signal strength than the 60$S_{1/2}$ spectrum, allowing for {higher measurement precision} than in the 60$S_{1/2}$ case.  Useful spectral features in the 58$D_{5/2}$ case again include the signal dropouts of several bands as well as points where the $m_j$~=~1/2 and 3/2 levels intersect for each sideband. For $D$ Rydberg levels of Rb, the low-field model based on a fixed polarizability~\cite{NJP_Adams} generally does not apply well due to the large variation of the polarizability $\alpha$ of these levels already at small fields (below 100~V/m in the present case). However, the band dropouts predicted by the non-perturbative Floquet model are accurate. More importantly, the hydrogenic lines that are clearly observed in the 58$D_{5/2}$ case provide excellent features with which to calibrate the field with high precision. Slight variations in alignment of the calculated spectra against the experimental ones lead to very noticeable variations in the achieved level of agreement. To elaborate on this point, in Fig.~\ref{figure_58D_triple}(c) we show an enlarged section of the hydrogenic manifold region of Fig.~\ref{figure_58D_triple}(a) with the results from Floquet calculation overlaid [Fig.~\ref{figure_58D_triple}(b) with a higher resolution]. We see excellent agreement in the overlap. While the beading of the experimental data, which is due to the experimental power step size, blurs most fine features, the avoided AC level crossings of some of the RF-modulated hydrogenic levels manifest as a corresponding variation in the separation between the spots observed in the experiment.

In the spectra, the bands are broadened in frequency to several tens of MHz due to saturation broadening of the probe transition, for which a higher-than-optimal intensity (about 11.5$~I_{\rm {sat}}$ with $~I_{\rm {sat}} = 3.9$~mW/cm$^2$~\cite{SteckRb85}) was used in order to achieve satisfactory signal levels to observe structures at the higher RF fields. In future work, this broadening could readily be reduced by heating the cell and reducing the probe power. Additionally, avoided crossings [such as those in the $m_j$=3/2 line at 16~dBi in Figs.~\ref{figure_58D_triple}(b)] could be better resolved. Unlike in previous measurements~\cite{TwoPhoton, HighPower}, in the present setup the inhomogeneity of the RF field has a negligible effect on the experimental line-widths.  This is because the diameter of the probe region ($70~\mu$m) is much smaller than the RF wavelength ($\ge~3$~m) and because the RF field does not vary significantly along the length of the cell (7.5~cm) which is smaller than the length of the parallel-plate capacitor [see Fig.~\ref{experimental_setup}(a)]. In future work one may also consider a reduction of the coupler Rabi frequency in order to reduce the weak-probe EIT bandwidth, $\Omega_c^2/\Gamma$ (which applies when $\Omega_p \lesssim \Omega_c \lesssim \Gamma$).

\section{DC offset}
\begin{figure}
\includegraphics[width=8.5cm]{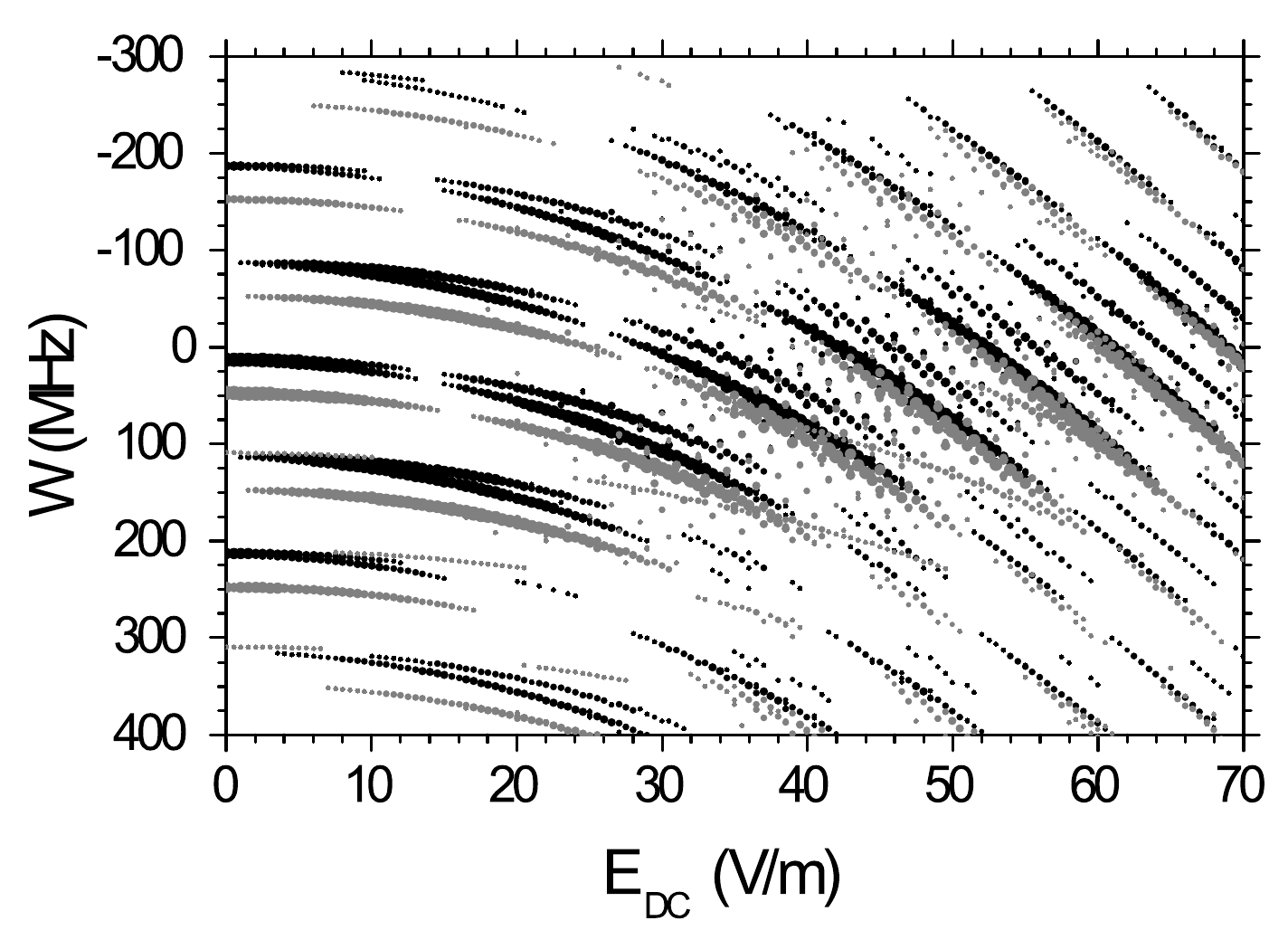}
\caption[caption]{Calculated spectrum of the 58$D_{5/2}$ state driven with 100~MHz RF with electric-field amplitude E$_{0,~RF}$~=~200~V/m as a function of applied DC field. Note that both even and odd sidebands are present. Black dots are for $m_j$~=~1/2 and gray dots are for $m_j$~=~3/2.}
\label{figureDC}
\end{figure}

In the above spectra only even harmonics are present, indicating the absence of any DC fields inside the spectroscopic cell.  Figure~\ref{figureDC} shows a calculated 58$D_{5/2}$ spectrum modulated with a 100~MHz RF field at a field amplitude of 200~V/m as a function of an additional applied DC electric field (the polarization of the DC field is the same as that of the AC field).  Here both even and odd harmonics are present. Based on this figure, for the 58$D_{5/2}$ state the lowest DC field that is resolvable is at about 1-2~V/m, which is when the odd sidebands begin to appear.

It has previously been suggested that DC fields external to the vapor cell are shielded by free charges within the cell. The free charges can originate from a photoelectric effect on the inside surface of the glass and/or from thermal or Penning ionization of Rydberg atoms~\cite{NJP_Adams}. This was investigated in our setup by applying a DC voltage across the two electrode plates, without application of an RF signal. For the 58$D$ and 82$D$ Rydberg-EIT spectra, measured for several probe-beam powers, there was no evidence of line broadening or shifting up to the experimental limit of the applied DC voltage, which was 1.2~kV. The corresponding DC test field of $\sim$400~V/cm (without cell) is about a factor of four higher than in a previous test~\cite{Adams_EIT}.

A lower limit of the DC shielding factor of the setup can be estimated based on the measured spectral line-width (several tens of MHz), the calculated DC polarizabity of the relevant D-states [several 100~MHz/(V/cm)$^2$ to several GHz/(V/cm)$^2$], and the applied DC field. In the present case, we estimate a shielding factor of $\sim 10^6$. The dielectric shielding due to the glass wall of the vapor cell, obtained from a field calculation, is estimated to be only $\sim$4$\%$. This result reiterates the ubiquitous finding that the interior volumes of spectroscopic alkali vapor cells are practically DC-field-free, even under exterior electrostatic boundary conditions that would suggest a strong interior DC field.

\section{Conclusion}
We have investigated the 60$S_{1/2}$ and 58$D_{5/2}$ Rydberg states in the presence of RF fields at 50~MHz and 100~MHz, respectively, using an all-optical readout method.  In moderate fields, we observe AC sidebands separated from the main line at even multiples of the driving frequency. At high fields, the modulated low-$\ell$ Rydberg levels intersect with a hydrogenic fan of high-$\ell$ states. The spectra in strong RF fields are modeled using Floquet theory. Using the 58$D_{5/2}$ spectrum, we achieved an electric-field measurement to within a $\pm$0.35$\%$ relative uncertainty. The spectral features that enable this level of uncertainty are signal dropouts of modulation bands at certain fields, intersections of low-$\ell$ Rydberg lines with steep levels that are part of the hydrogenic fan of states, and level crossings and narrow anti-crossings in the spectra. The determination of the RF electric field was based on an atomic response and did not require calibration of RF components. We found a shielding factor of $\sim 10^6$ for intentionally applied DC offset fields, which agrees with previous findings elsewhere. The spectroscopic RF-field measurement method is therefore largely immune to accidentally occurring external static fields.

This work is supported by NSF Grant Nos. PHY-1205559 and PHY-1506093.

\end{document}